%% file: main.tex
\documentclass[manuscript,screen]{acmart}

\usepackage{multirow}

\AtBeginDocument{%
  \providecommand\BibTeX{{%
    \normalfont B\kern-0.5em{\scshape i\kern-0.25em b}\kern-0.8em\TeX}}}

\setcopyright{acmcopyright}
\copyrightyear{2018}
\acmYear{2018}
\acmDOI{XXXXXXX.XXXXXXX}

\acmConference[IUI '24]{Make sure to enter the correct
  conference title from your rights confirmation emai}{March 18--21,
  2024}{Greenville, SC}
\acmBooktitle{IUI '24: ACM Conference on Intelligent User Interfaces (ACM IUI),
 March 18--21, 2024, Greenville, SC} 
\acmPrice{15.00}
\acmISBN{978-1-4503-XXXX-X/18/06}

\usepackage[colorinlistoftodos]{todonotes}

\copyrightyear{2024}
\acmYear{2024}
\setcopyright{acmlicensed}\acmConference[IUI '24]{29th International Conference on Intelligent User Interfaces}{March 18--21, 2024}{Greenville, SC, USA}
\acmBooktitle{29th International Conference on Intelligent User Interfaces (IUI '24), March 18--21, 2024, Greenville, SC, USA}
\acmDOI{10.1145/3640543.3645175}
\acmISBN{979-8-4007-0508-3/24/03}

\begin{document}

\title{Chart4Blind: An Intelligent Interface for Chart Accessibility Conversion}


\input{chapters/auth}
\renewcommand{\shortauthors}{Moured and Baumgarten-Egemole et al.}


\input{chapters/abstract}

\input{chapters/teaser}



\maketitle


\input{chapters/introduction}


\input{chapters/relatedwork}


\input{chapters/needfinding}


\input{chapters/chart4blind.tex}


\input{chapters/userstudy}


\input{chapters/evaluatingoutputaccessibility}


\input{chapters/limitationsandfuturework}


\input{chapters/conclusion}


\begin{acks}
This work was supported by the European Union’s Horizon $2020$ research and innovation program under the Marie Sklodowska-Curie Grant No.$861166$. The AI experiments were conducted in part on the HoreKa supercomputer, funded by the MWK and the Federal Ministry of Education and Research.
\end{acks}


\bibliographystyle{ACM-Reference-Format}
\bibliography{main}

\appendix

\end{document}

%% file: chapters/auth.tex

 \author{Morris Baumgarten-Egemole}
 \authornotemark[1]
 \email{morris.baumgarten-egemole@student.kit.edu}
 \orcid{}
 \affiliation{%
   \institution{ACCESS@KIT, Karlsruhe Institute of Technology}
   \streetaddress{Adenauerring (Geb. 50.28)}
   \city{Karlsruhe}
   \state{Baden-Württemberg}
   \country{Germany}
   \postcode{76131}
 }
 
 \author{Omar Moured}
 \authornotemark[1]
 \email{omar.moured@kit.edu}
 \orcid{0000-0003-4227-8417}
 \affiliation{%
   \institution{CVHCI \& ACCESS@KIT, Karlsruhe Institute of Technology}
   \streetaddress{Adenauerring (Geb. 50.28)}
   \city{Karlsruhe}
   \state{Baden-Württemberg}
   \country{Germany}
   \postcode{76131}
 }

\author{Karin Müller}
 \orcid{0000-0003-4309-1822}
 \email{karin.e.mueller@kit.edu}
 \affiliation{%
   \institution{ACCESS@KIT, Karlsruhe Institute of Technology}
   \streetaddress{Adenauerring (Geb. 50.28)}
   \city{Karlsruhe}
   \state{Baden-Württemberg}
   \country{Germany}
   \postcode{76131}
 }

 \author{Alina Roitberg}
 \email{alina.roitberg@f05.uni-stuttgart.de}
 \orcid{0000-0003-4724-9164}
 \affiliation{%
   \institution{Institut für Künstliche Intelligenz, Universität Stuttgart}
   \streetaddress{Adenauerring (Geb. 50.28)}
   \city{Karlsruhe}
   \state{Baden-Württemberg}
   \country{Germany}
   \postcode{76131}
 }

 \author{Thorsten Schwarz}
 \orcid{0000-0002-7346-5744}
 \email{thorsten.schwarz@kit.edu}
 \affiliation{%
   \institution{ACCESS@KIT, Karlsruhe Institute of Technology}
   \streetaddress{Adenauerring (Geb. 50.28)}
   \city{Karlsruhe}
   \state{Baden-Württemberg}
   \country{Germany}
   \postcode{76131}
 }

 \author{Rainer Stiefelhagen}
 \email{rainer.stiefelhagen@kit.edu}
 \orcid{0000-0001-8046-4945}
 \affiliation{%
   \institution{CVHCI \& ACCESS@KIT, Karlsruhe Institute of Technology}
   \streetaddress{Adenauerring (Geb. 50.28)}
   \city{Karlsruhe}
   \state{Baden-Württemberg}
   \country{Germany}
   \postcode{76131}
 }

\authornote{Both authors contributed equally to the paper}


%% file: chapters/abstract.tex
\begin{abstract}
In a world driven by data visualization, ensuring the inclusive accessibility of charts for Blind and Visually Impaired (BVI) individuals remains a significant challenge. Charts are usually presented as raster graphics without textual and visual metadata needed for an equivalent exploration experience for BVI people. Additionally, converting these charts into accessible formats requires considerable effort from sighted individuals. Digitizing charts with metadata extraction is just one aspect of the issue; transforming it into accessible modalities, such as tactile graphics, presents another difficulty. To address these disparities, we propose \textit{Chart4Blind}, an intelligent user interface that converts bitmap image representations of line charts into universally accessible formats. Chart4Blind achieves this transformation by generating Scalable Vector Graphics (SVG), Comma-Separated Values (CSV), and alternative text exports, all comply with established accessibility standards. Through interviews and a formal user study, we demonstrate that even inexperienced sighted users can make charts accessible in an average of $4$ minutes using \textit{Chart4Blind}, achieving a System Usability Scale rating of $90\%$.  In comparison to existing approaches, \textit{Chart4Blind} provides a comprehensive solution, generating end-to-end accessible SVGs suitable for assistive technologies such as embossed prints (papers and laser cut), 2D tactile displays, and screen readers. For additional information, including open-source codes and demos, please visit our project page \href{https://moured.github.io/chart4blind/}{https://moured.github.io/chart4blind/}.

\end{abstract}

\keywords{Data Visualization, Chart, Digitization, Accessible Charts, Embossed Paper}

\ccsdesc[500]{Human-centered computing~Accessibility systems and tools}

%% file: chapters/teaser.tex
\begin{teaserfigure}
  \includegraphics[width=\textwidth]{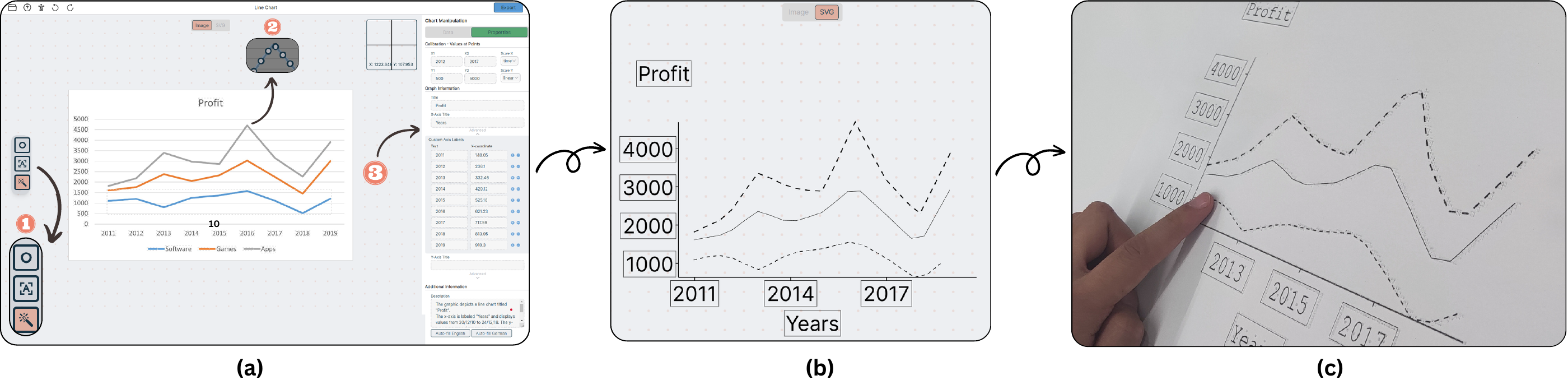}
  \caption{Using the Chart4Blind interface, a sighted person can upload a bitmap image, which is automatically analyzed, and accessibility relevant information is inserted. The interface shown in \textbf{(a)} allows manual adaptations to improve accessibility. The export function translates the recognized text to Braille fonts and lines to lines, that can be embossed and distinguished tactilely \textbf{(b)}. A sample outcome of the export is an embossed tactile graphic with a short textual description for finger exploration by people with visual impairments \textbf{(c)}.}
  \Description{The picture depicts three subfigures. In (a), the Chart4Blind interface is illustrated, highlighting three AI tools and the metadata fields. A simple line chart with three lines is also visible in the UI. (b) displays the same line chart as in (a), but in print-accessible mode. (c) showcases the printed version of the chart, indicating the complete flow of the chart in the conversion process.}
  \label{fig:teaser}
\end{teaserfigure}

%% file: chapters/introduction.tex
\section{Introduction}





Humans heavily rely on graphs and charts to analyze trends, make comparisons, and extract meaningful insights in both academic and commercial areas. In many professions, an effective understanding of graphical data is a crucial prerequisite for achieving success. The rapidly growing amount of information available online~\cite{Tylor2023} has shifted the focus of research from gathering data to ensuring it is accessible and understandable to all. Line graphics are a widespread way of presenting information, e.g., when teaching students, presenting a new idea to business partners, or exchanging knowledge in general. 

However, most existing chart representation methods are predominantly \textit{visual}, posing significant accessibility challenges for people with visual impairments. 
Despite the advancements in data visualization tools, there remains a significant gap in ensuring that these tools are inclusive and accessible for all.  \textit{Line charts} used in books, learning materials, and on slides are a common type of graph. Typically, these documents lack textual and visual metadata, which prevents people with visual impairments from exploring the graphics.
 Moreover, the process of converting these charts into accessible formats such as textual descriptions or tactile graphics is often labor-intensive and requires significant effort from sighted individuals~\cite{sheppard2001tactile}.
Although tools to automatically describe graphs textually have been more explored in recent years,  ~\cite{Lundgard_2022,Sharif-etal_2022,tang2023vistext}, approaches to automatically extract information from graphs and generate tactile graphics have hardly been addressed in research.
While a handful of studies have approached this from a computer vision standpoint~\cite{SINGH2023100555, moured2023accessible, moured2023line}, the user experience aspect of people who create accessible graphics has been largely overlooked and is the main motivation of this work. 

Our main goal is to create a tool that helps sighted people make graphs accessible to people with visual impairments.
To address this goal, we introduce \textit{Chart4Blind} -- an intelligent user interface designed to transform bitmap image representations of line charts into universally accessible formats \cite{nagao2005braille, edman1992tactile}, including SVG, CSV files, and alternative text. These formats can be used to emboss a tactile graphic on paper or to read it on a 2D haptic Braille display.
While our design principles primarily target the visually impaired and blind individuals, our framework has strong potential for saving time of a sighted user e.g. assistants, friends, or family members of blind persons by supporting automatic digitization and information extraction from line graphs. \textit{Chart4Blind} ensures that the converted charts align with established accessibility standards \cite{w3c} and supports the process of transfer of line charts. We started with a series of \textit{need-finding interviews} to guide our design choices. The tool design emphasizes clear progress indicators during conversion, consistency in the conversion steps, intuitive data input methods, and integration of AI tools for efficient data extraction.
To validate our tool, we conducted a usability study with 10 sighted participants aged 22-34 years.
The study revealed that the \textit{Chart4Blind} tool is valuable for converting line charts into accessible formats and helps sighted users understand accessibility guidelines with an average System Usability Scale rating of 90\%. To validate the accessibility of the produced graphic for BVI in particular, we conducted a follow-up study with 3 blind individuals, asking open-ended questions regarding the output of our interface and collecting valuable feedback for future improvement.

This paper introduces Chart4Blind, an intelligent user interface designed to convert bitmap image representations of line charts into universally accessible formats, and has the following major contributions:

\begin{itemize}

    \item \textbf{Chart4Blind System:} Through a series of interviews and feedback, we designed an intuitive tool for converting bitmap line charts into multiple universally accessible formats. It also supports collaborative efforts where multiple people can work together and assist in the conversion process. \textit{Chart4Blind} has a user-friendly interface and can be utilized by individuals with varying levels of experience.

    \item \textbf{Integration of Intelligent Features:} The tool incorporates intelligent deep learning models to ensure a seamless conversion process, particularly OCR and segmentation models. Users can interact with the model predictions through simplified actions, such as drag-and-drop.
    
    \item \textbf{Usability Study:} A thorough usability study involving sighted people aged between 22-34 years to validate the effectiveness of the \textit{Chart4Blind} system. Furthermore, another user study was conducted specifically with blind participants to assess the output quality and accessibility. The tool achieved an average System Usability Scale rating of $90\%$.
\end{itemize}
 


%% file: chapters/relatedwork.tex
\section{Related Work}
Our research builds upon the fundamental elements of earlier studies, which include: (1) accessible charts, (2) chart deconstruction, and (3) chart summarization.

\subsection{Accessible Representation of Charts}
People with visual impairments often struggle with bitmap images of charts~\cite{10.1145/1168987.1169018, SINGH2023100555}. 
To address this, the Web Content Accessibility Guidelines recommend offering a textual description of the chart alongside its graphical representation as alternative text\cite{world1999web}. 
Yet, such descriptions are rarely created by authors \cite{10.1145/1168987.1169018}. To reduce the manual effort in this task, some works automate alternative text generation \cite{balaji2018chart, kantharaj2022charttotext, 10192564}. 
Unfortunately, such tools have several issues, such as producing irrelevant information and hallucinations~\cite{tang2023vistext, kantharaj2022charttotext}.

While the alternative text is useful for describing line charts, it is not always enough for those with visual impairments \cite{altmanninger2006dynamically, watanabe2012development}.  There are three alternative ways to present graphical information for visually impaired people: (1) Tactile graphics, using relief elements for haptic perception \cite{serrano2021study}, (2) Alternative Text, describing graphical content in words \cite{devine2011making}, or with screen readers \cite{zou2015chartmaster, blanco2022olli} and (3) Sonification, mapping raw data values to a diverse range of sounds, varying pitch, frequency, and tone to enable easy distinction between line trends \cite{holloway2022infosonics, bru2023line,hermann2011sonification}.

When it comes to tactile graphics, they can be in different formats: embossed paper, swell paper, thermoform, laser cut, and 3D printed~\cite{Butler-etal_2021}. These formats typically provide better dot resolutions, utilizing various pin height levels to represent more information, and they're cost-effective and portable for individuals with visual impairments \cite{li2019editing, engel2018user}. However, they lack the capability for advanced interaction. The second format is digital tactile displays, which can be refreshed and offer additional features such as zooming, interaction buttons, and audio output \cite{moured2023accessible}. Nonetheless, this option tends to be more expensive and offers lower pin resolution. 
In our work, we ensure that the converted charts meet the accessibility needs of both digital and printed formats.

Having access to a chart's raw data, which includes plot data, titles, axis labels, and descriptions, is highly beneficial. For example, some works use this data to create tactile graphics like Audio-Tactile charts, which combine touch interaction with audio feedback \cite{10.1145/2661334.2661389, engel2019svgplott}. Similarly, Sonification and alternative text descriptions benefit from raw data access.
In terms of raw data extraction methods, they can be categorized as: (1) Manual, requiring human intervention without automated tools, like Data Thief \cite{Tummers2006}, (2) Semi-automatic, combining automatic features with some human intervention \cite{rohatgi2014webplotdigitizer, 10.1145/3025453.3025957, vaingast2014im2graph, 10.1145/3544548.3581113, Huwaldt2023, Tummers2006}, and (3) Fully automatic, with no manual intervention \cite{lal2023lineformer,choi2019visualizing}, though they have limitations in the conversion accuracy.

Given the importance of raw data, vector graphics, such as SVGs, offer a promising alternative representation \cite{godfrey2014statistical}. SVGs, created using a chart's raw data, have features that make them more accessible than bitmap images \cite{ferraiolo2000scalable}. Their structure supports tactile printable graphics creation \cite{10.1145/1090785.1090816}. For instance, Braille printers can emboss raised dots to mark outlines \cite{10.1145/1090785.1090816}. Additionally, methods like LineSpace \cite{10.1145/2858036.2858245}, which use 3D printer filament to print SVG file elements, show the potential of SVGs. In conclusion, SVGs can complement traditional alternative text descriptions for line charts.

\subsection{Image to Vector Graphic Conversion}
Having determined SVG as a suitable format for an alternative representation, we now describe techniques for converting bitmap images to vector graphics.  
Tools such as Libre Office Draw can be used to convert images to a Libre Office vector graphic format~\cite{libreofficedraw}, but this method is time-consuming~\cite{prescher2014production}. 
To address these challenges, Jayant et al. introduced an automated solution that converts bitmap images into Adobe Illustrator vector graphics \cite{10.1145/1296843.1296858}. 
However, this tool requires manual training on similar charts and directly translates charts into printable graphics, which is not ideal for tactile charts due to braille embosser constraints \cite{engel2017improve}.
Several guidelines have been introduced to govern the conversion process \cite{gale2005guidelines, gale2022guidelines}. 
Goncu et al. proposed a tool that converts pie and bar chart data into SVG \cite{10.1145/1414471.1414525}, arguing against a one-size-fits-all approach for accessible chart representation. Offering multiple output options can diminish barriers between BVIP and sighted users \cite{10.1145/1414471.1414525}, making the original and tactile charts more alike, potentially facilitating BVI's interpretation of chart data \cite{engel2018user}.

\subsection{User Experience Analysis}
Usability is vital when designing interactive user interfaces \cite{unger2012project}.
Recent applications emphasize intuitive design. Tools such as ChartDetective~\cite{10.1145/3544548.3581113} and PlotDigitizer~\cite{Huwaldt2023} are web-apps that adhere to design best practices, incorporating Visual Information-Seeking Mantra principles~\cite{shneiderman1996eyes}. However, the former tool, ChartDetective, accepts only vector graphic charts in PDF format, and neither tool supports accessible output formats.

Enhancing UX also involves offering a magnified view around the mouse pointer, improving application accuracy. 
While tools like im2graph~\cite{vaingast2014im2graph} and WebPlotDigitizer \cite{rohatgi2014webplotdigitizer} emphasize such explorations.

From 2006 to 2023, across various chart analysis tools \cite{Tummers2006, Huwaldt2023}, manual calibration of chart axes, involving setting four calibration points to map pixel values to the x-y plane, has remained a consistent feature. Despite the potential benefits of automation, its implementation is limited by current algorithms, which only achieve 61.7\% accuracy in axis detection \cite{moured2023line}. Thus, manual methods, either by prompt-based clicking \cite{Tummers2006, rohatgi2014webplotdigitizer} or drag-and-drop \cite{Huwaldt2023}, are preferable.
Semi-automation also aids in text value entry. Im2Graph~\cite{vaingast2014im2graph} and ChartDetective~\cite{10.1145/3544548.3581113} employ OCR to recognize image text \cite{chaudhuri2017optical, georgenagy}. After calibration, various techniques exist for extracting plot data, often resulting in CSV files \cite{rohatgi2014webplotdigitizer, 10.1145/3025453.3025957, vaingast2014im2graph, 10.1145/3544548.3581113, Huwaldt2023, Tummers2006, Huwaldt2023}. WebPlotDigitzer and PlotDigitizer offer semi-automatic extraction, letting users edit detected data markers. Im2graph employs a color-based line detection approach. Data markers, which can mimic curves and match exported values, are commonly used due to their predictable results.

%% file: chapters/needfinding.tex
\section{NEED-FINDING INTERVIEWS}
To gain a comprehensive understanding of how sighted individuals approach the creation of accessible line charts for BVI individuals, and to identify how future tools can efficiently support their efforts, we conducted a series of exploratory need-finding semi-structured interviews. While recent research has focused on the challenges involved in extracting chart metadata \cite{cheng2023chartreader, lal2023lineformer, hassan2023lineex}, it has not thoroughly addressed the specific requirements of end users, particularly the blind and visually impaired. Furthermore, there has been a gap in exploring the essential prerequisites for an online tool to streamline this conversion process. Therefore, our need-finding interviews were carried out with the primary objective of uncovering these specific design requirements. 
Our inquiries primarily aimed to:
\begin{itemize}
    \item Understand the limitations of common interfaces and user practices for converting charts into accessible formats.
    \item  Pinpoint the features that users find important in such a tool.
    \item  Gain insights into how AI tools are explored in prior research and could facilitate the creation of accessible charts.
\end{itemize}
\subsection{Participants}
We conducted a semi-structured interview with four sighted participants (P1-4, 2 female, and 2 male, age range 25-40 years) who would use the Chart4Blind user interface. We collaborated with a local non-profit accessibility service center ACCESS@KIT\footnote{https://www.access.kit.edu/}, to identify experts who transfer literature for blind users. All sighted individuals exhibited proficiency in working with line charts, familiarity with accessibility guidelines, and practical experience in converting chart materials. Three of the participants occasionally converted charts for BVI students, while one participant undertook this task on a daily basis. Table \ref{tab:nfp} illustrates the working domains and experience level of our participants in more detail. The participants received no compensation. The study was part of a series of studies which were approved by the ethical review committee of Karlsruhe Institute of Technology.

\begin{table}
  \caption{Participants' overview: Experience in BVI chart conversion and approximate conversion frequency.}
  \label{tab:nfp}
  \begin{tabular}{c|c|c|c|c|l}
    \toprule
    Participant & Gender & Field & Experience & Frequency & Tool used \\
    \midrule
    P1 & F & Assistive Technologies & +5 years & Daily basis & LibreOffice Draw \cite{libreoffice}
    \\
    P2 & F & Computer Science & +2 years & once every 2 weeks & LibreOffice Draw \\
    P3 & M & Computer Science & +10 years & once a month & Inkscape \cite{inkscape}\\
    P4 & M & Physics & +2 years & once a month & WebPlotDigitizer \cite{rohatgi2014webplotdigitizer}\\
  \bottomrule
\end{tabular}
\end{table}
\subsection{Procedure}
During our need-finding study, we asked participants open-ended questions about their experiences, challenges, and requirements for the chart conversion process. We asked the following questions:
\begin{verbatim}
   (1) Describe your workflow for converting charts into an accessible format. 
   (2) What challenges did you encounter during this process?
   (3) What specific computer-based tools do you utilize for the conversion process?
\end{verbatim}
We followed up by inquiring about the steps they found most challenging and time-consuming. Depending on the interview, we additionally asked: 
\begin{verbatim}
    (4) Could you estimate the time required for chart conversion?
    (5) Which features of your current tools do you find most useful?
\end{verbatim}

These one-to-one interviews were conducted face-to-face, with the interviewer asking questions and taking written notes, and they typically lasted around 40-45 minutes. After reviewing the interview notes for mentioned topics, we established the design principles and requirements. This led to the development of the \textit{Chart4Blind} prototype using Figma\footnote{https://www.figma.com/}, a popular design tool widely used by user interface designers. During this phase, participants P1 and P3 assisted in refining minor aspects of Chart4Blind such as color schemes, font styles, and the tutorial script.
\subsection{Findings}
\label{findings}
In this section, we present the primary themes derived from our analysis, as well as the identified user requirements (UR) that will guide our \textit{Chart4Blind} prototype design.

\subsubsection{Needs for streamlined process} Our participants employed distinct tools for the conversion task. P1-3 utilized drawing tools \cite{inkscape, libreoffice} to overlay components onto chart images, replicating the original chart's elements. On the other hand, P4 employed WebPlotDigitizer \cite{rohatgi2014webplotdigitizer} to extract metadata and then used data visualization tools to regenerate an accessible chart. When asked about the average time required for a single chart conversion, participants agreed that it depends on the chart's complexity. None reported taking less than 10 minutes, even for easy charts, as the process requires manual drawing, adding text labels, and writing a chart description.

Despite the diverse tools used, we identified common steps in the conversion process, highlighting an opportunity to streamline this task into a unified tool. However, participants encountered challenges involving inexperienced individuals for assistance, as the existing tools were not designed for this purpose. Extensive tutorials were necessary to prepare individuals for the task. Additionally, the process lacked the potential for parallelization due to limitations in (real-time) data-sharing mechanisms within the current tools. P1 emphasized how these challenges prevented effective crowdsourcing efforts. In response to these findings, we propose developing a user interface that caters to both experts and non-experts, presenting clear and consistent steps (UR-1). Furthermore, the tool facilitates a more parallelizable process to encourage concurrent contributions (UR-2).

\subsubsection{Needs for intelligent features} The analysis of question 4 highlighted two time-consuming tasks: text extraction and line segmentation, aligning with charts being both visual and textual. P4 mentioned that many charts have numerous axis ticks, often requiring manual rewriting. P1-2 discussed line segmentation challenges, especially with more lines or complex trends like sinusoidal patterns. P3, using WebPlotDigitizer, noted the tool's reliance on color filters for segmentation, requiring parameter tuning. These insights highlight the need for a robust OCR (UR-3) and a deep learning-based line annotation tool (UR-4). Recognizing diverse metadata needs for various charts, we opted for a semi-automatic approach over an end-to-end solution.

\subsubsection{Needs for accessible output} Another aspect missing in other tools is ensuring an accessible export. Most participants (P1-3) prefer drawing tools due to their SVG export capability. However, two challenges were highlighted. Firstly, P3 pointed out that the drawing approach did not include line metadata in the export, containing only the line drawing position relative to canvas dimensions and not the real coordinates. Secondly, P1 mentioned the need for further processing on the export, especially for complex charts that require simplification for BVI individuals such as reducing the number of included axis labels to fit the embossed paper size or optimizing the SVG for screen readers by removing unnecessary tags and adding descriptive ones.

To our knowledge, no tool currently enables metadata extraction and produces a printable export suitable for BVI individuals. Given varying needs, sometimes leaning towards metadata or visual elements, we introduced the requirement to support accessible extraction in both CSV and SVG formats (UR-5). The decision was informed by the support of other BVI assistive technologies for these formats \cite{engel2019svgplott, thompson2023chart}.
\subsection{Design Principles}
Based on our interview findings, we have identified key design principles that guide our \textit{Chart4Blind} system:

\begin{itemize}
    \item \textbf{Clear Progress Indication:} The tool should provide users with a clear view of their current progress during chart conversion, indicating what has been completed and what remains.
    
    \item \textbf{Maintain Consistent Steps:} The tool should ensure that the steps in the conversion process remain consistent regardless of chart complexity, allowing for a more parallelized process. 

    \item \textbf{Intuitive Data Input:} The tool should allow for an intuitive approach to make sure that  all necessary information needed for an accessible chart is added. In addition, the tool should also support intuitive interactions such as drag-and-drop for textual content and drop-down lists for axis label formats (e.g., linear, logarithmic).
    
    \item \textbf{Automated Selection Tools:} To reduce the efforts spent extracting metadata, the tool should integrate an automated solution to ease the task of element selection (e.g., lines, texts). In our case, we choose AI-driven tools for line segmentation, text extraction, and chart description when needed. 
    
    \item \textbf{Ready-to-Use Accessible Exports:} The tool should provide users with accessible output for various modalities such as tactile printing and digital displays used with a screen reader.

\end{itemize}

%% file: chapters/chart4blind.tex
\section{The Chart4Blind System}
In this section, we describe the design of our Chart4Blind system, which assists sighted people in making charts accessible for people with visual impairments by following the design principles we derived from our need-finding interviews.
Our system comprises two main modules: (1) Data Extraction Module and (2) Rendering Module. 

\begin{figure}[h]
  \centering
  \includegraphics[width=\linewidth]{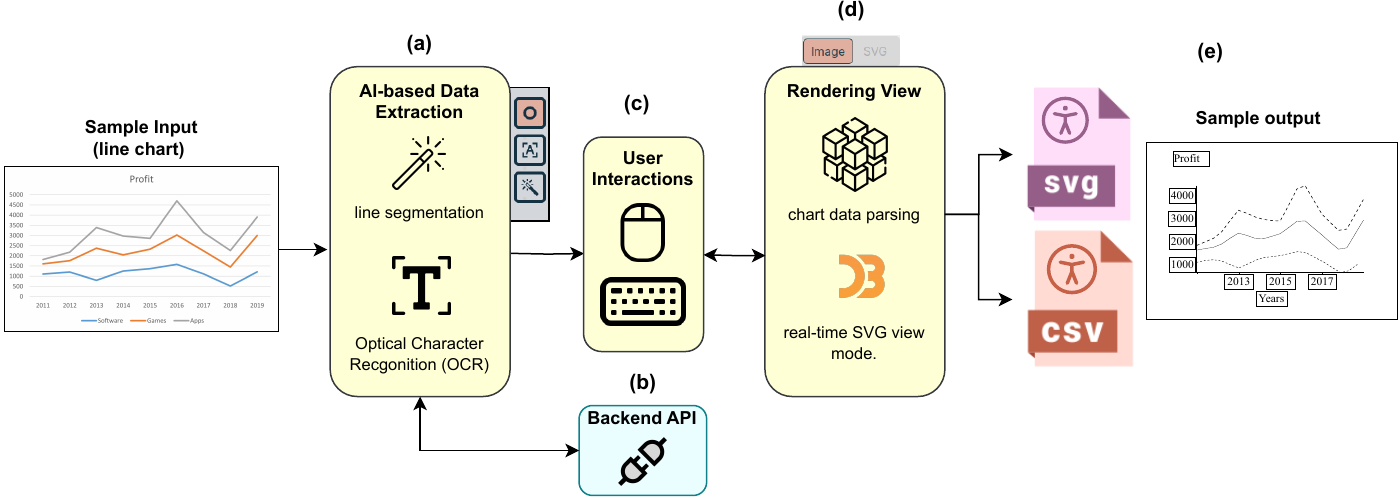}
  \caption{The pipeline of Chart4Blind consists of the input of a bitmap line chart, followed by the Data Extraction Module, which includes an AI-based line segmentation and optical character recognition step, and a manual correction step by a sighted user (a). The Rendering Module updates the information in real-time and ensures an accessible representation (c). The system allows the export of the information to an SVG and a CSV format. The SVG can be accessed with a screen reader or printed as a tactile graphic (d). The metadata can be exported as accessible CSV as well.}
  \Description{The picture illustrates the pipeline of Chart4Blind. On the far left, a sample input line chart with three lines enters the first stage (a), which involves AI-based data extraction. Here, the user can interact with the model predictions (c). The user also has the option to view progress using different view modes through the rendering view options (d). Finally, the outputs can be exported, as shown, to SVG or CSV, and a sample output is depicted on the far right.}
  \label{fig:chart4blindsystem}
\end{figure}

Figure \ref{fig:chart4blindsystem} shows the pipeline of \textit{Chart4Blind} system. When a user uploads a line chart image, a toolbar appears that is linked to our AI-based \textit{Data Extraction Module}, which is responsible for extracting the chart's textual and visual data. After calibrating four axis points, this intelligent module converts pixel values into the raw data domain. Subsequently, our SVG viewing mode, connected to the \textit{Rendering Module}, generates the SVG view in real-time as users update field information. All session information is stored upon consent acceptance, and a unique Token is assigned to facilitate collaborative work. The final step includes the export of the visual data to a printable tactile graphic format (SVG) and to a textual description (CSV). The following paragraphs describe the two main modules in more detail.

\subsection{Data Extraction Module}
Charts typically encompass two types of information: textual and visual. For each type, we incorporate an intelligent tool to facilitate extraction with minimal user interactions. In order to accomplish this, we conducted a comprehensive review of the state-of-the-art in this field \cite{bajic2023review, masson2023chartdetective, ying2023reviving} and opted to utilize a segmentation model for line data extraction and OCR for the textual content. The reason behind our selection can be summarized as follows:

\begin{itemize}
    \item A light model capable of running offline on standard consumer hardware avoiding external server infrastructure and  minimizing backend computational demands.

    \item A model that generates results that could be adjusted and validated by the user in order to simplify the interaction. 
\end{itemize}

\subsubsection{Instance Segmentation Feature}
A line chart primarily exists in the pixel domain. The image itself does not contain the original table data used for its creation. Hence, a crucial step involves converting from the pixel domain to the raw domain. This conversion task requires three specific conditions. Firstly, segmentation of the line of interest. Secondly, two calibration points on each axis to map pixels, given that the axes are straight lines. Lastly, the data format of labels (e.g., linear, logarithmic, date, etc.).

To address the initial challenge with minimal interactions, we conducted experiments utilizing instance segmentation models. This approach enables the segmentation of each line into a distinct mask, providing the flexibility to trace each line. Users gain control over selecting which line to segment. Motivated by Ying et al. work \cite{lal2023lineformer}, we trained a Mask2Former \cite{cheng2021mask2former} on the \textit{LG} dataset \cite{moured2023line}. Unlike \textit{LineFormer}, the \textit{LG} dataset we used comprises real charts with human annotations, which we found to be more robust for the charts users might upload. The model achieved a score of 62.12 IOU (Intersection over Union) on the test set. From the segmentation masks, we sampled key points, equidistant from each other, as demonstrated in Figure \ref{fig:interface}. Overlaying the masks on the chart empowers the user to choose which line to process and allows control over increasing or decreasing the number of key points extracted.

Regarding the calibration points, we offer users four initialized points positioned close to the Y and X axes. Users are then guided to position these four calibration points onto the axis ticks, input the corresponding label values, and choose the data format from a drop-down menu, including time, logarithmic, and linear data types. The conversion to the raw domain is then carried out following the same linear approach as \cite{rohatgi2014webplotdigitizer} and \cite{Huwaldt2023}.

\subsubsection{OCR Feature}
To categorize the textual content in the line chart, we followed the analysis outlined in \cite{tang2023vistext} and \cite{moured2023line}. The chosen categories are shown in Table \ref{tab:textfields}. For each category, a corresponding text field has been added within the metadata tab in the interface. We utilize the Tesseract OCR system \cite{kay2007tesseract}, operating on the user's browser even when offline. This particular model has demonstrated a good performance in the ICDAR chart text understanding challenge \cite{yao2015incidental}. We employed a template-based approach \cite{alam2023seechart} to auto-fill chart descriptions. This equipped the chart with a summary of the encoded elements and descriptive statistics (e.g., extremes, outliers, etc.), corresponding to Levels 1 and 2, as proven preferable by many BVI individuals \cite{lundgard2021accessible}.

\begin{table}
  \caption{Text fields present in the Chart4Blind interface.}
  \label{tab:textfields}
  \begin{tabular}{ccl}
    \toprule
    Property & Fields \\
    \midrule
    Calibration & Axes calibration points\\
    \cmidrule{2-2}
    \multirow{3}{*}{Chart Information}  & Plot title \\
      & Axes titles \\
      & Axes labels \\
    \cmidrule{2-2}
    \multirow{2}{*}{Additional Information} & Chart description  \\
       & Data point description  \\
  \bottomrule
\end{tabular}
\end{table}
\subsection{Rendering Module}
To address UR-1 and UR-5 findings, we incorporated a real-time view to track conversion progress. Users can switch between modes to visualize textual or line drawings. Our rendering module displays a real-time SVG view using D3.js and exports results at the end of the session. D3.js was chosen for its memory efficiency and rich feature set, surpassing other DOM manipulation methods \cite{benbba2021comparison}. It also complies with W3C \cite{w3c} standards and enables interactive chart creation, beneficial for future audio integration to test screen reader accessibility.

Considering the space and size constraints of printed charts on embossed paper, adherence to several print guidelines for tactile illustrations \cite{BANA2010, nagao2005braille, edman1992tactile} is essential. Related requirements are summarized as follows:
\begin{itemize}
    \item Lines should be capable of being distinguished by touch, either by using different thicknesses or different types of symbols such as a dotted or dashed line.
    \item  Lines of < 0.4 mm thick should not be used, as it can be difficult to obtain a sufficient bump on capsule paper.
    \item Text in tactile illustrations should be written in Braille and oriented horizontally. A margin of at least 3.0 mm should be left around the Braille characters.
\end{itemize}

\begin{figure}[h]
  \centering
  \includegraphics[width=0.9\linewidth]{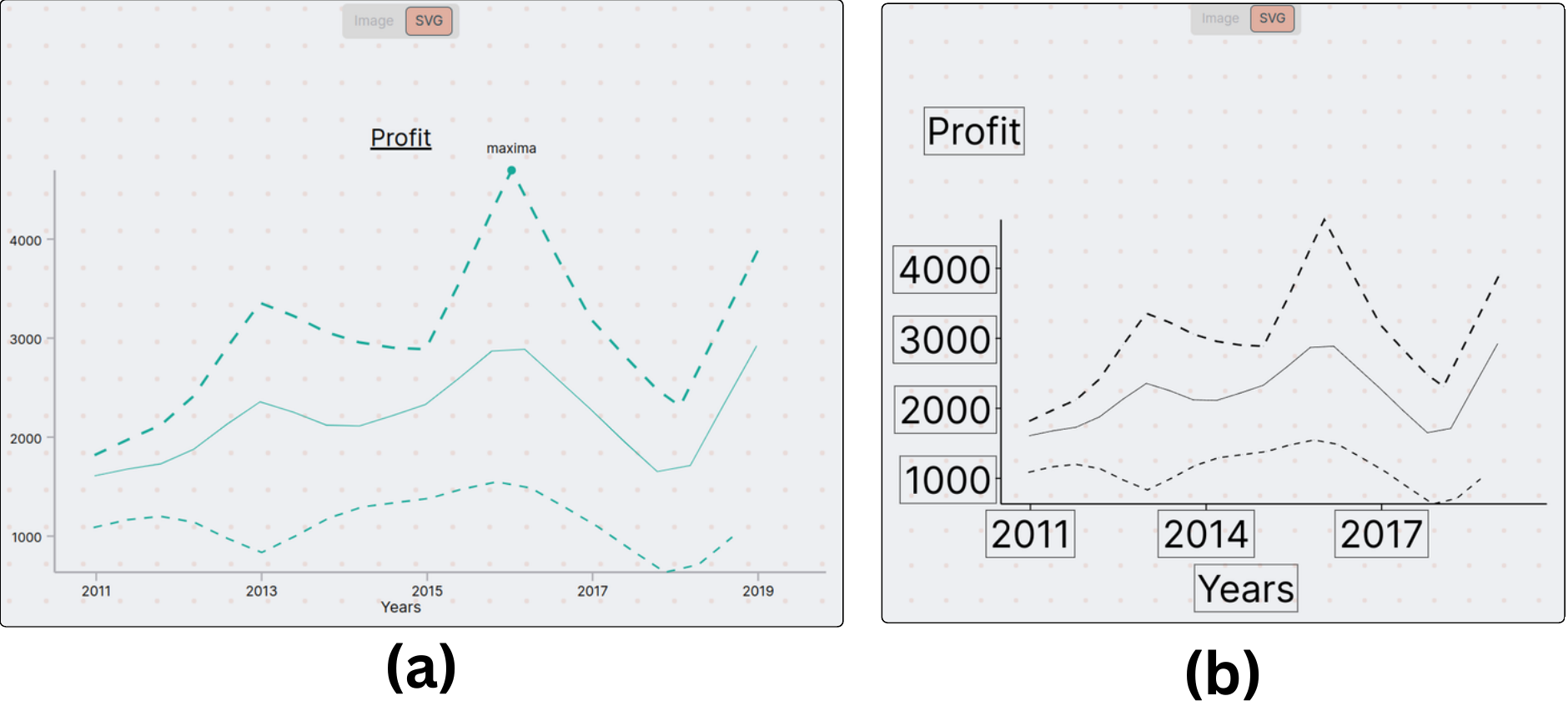}
  \caption{Rendering Module view for a bitmap line chart. (a) displays the digitally accessible SVG view, ideal for screen readers, and refreshable tactile devices. (b) shows the print-accessible SVG view, suitable for print modalities such as embossed papers or laser cut.}
  \Description{The image exhibits two screenshots from the UI Rendering Module. On the left, the SVG is finely detailed and suited for screen readers or technologies that provide enhanced interaction modalities. On the right, a more coarse representation of the chart to accommodate the necessary information in the printed version, such as larger text.}
  \label{fig:accessiblemodes}
\end{figure}

Our rendering module offers an additionally accessible visualization mode, allowing users to export SVGs that align with printing guidelines for tactile graphics~\cite{BANA2010}. This includes a reduced number of axes labels, typically limited to 3 to 5 labels depending on the page size due to the size of Braille script.  The (default) digitally accessible SVG mode is more suitable for 2-D tactile displays and screen readers (Figure \ref{fig:accessiblemodes}-(b)), enabling the embedding of more visual content with corresponding description tags (e.g. <desc>). Figure \ref{fig:accessiblemodes} illustrates a sample rendering outputs for both digital and print-accessible SVG modes. 

Although line charts come in various styles and layouts, our rendering is well-suited for univariate charts with only one type of data and a single set of x and y coordinates. Additionally, since our segmentation model doesn't automatically predict line encodings, users have the option to set them manually as needed.

\subsection{Chart4Blind Interface}
Informed by insights obtained from need-finding interviews with our participants and guided by our design principles, we developed the Chart4Blind interface. This interface allows users to upload a chart image and facilitates the generation of an accessible version with text descriptions in diverse formats. In the course of our design process, we engaged two participants from the need-finding interviews to gather expert feedback. We carefully considered this feedback to enhance the seamless conversion process within the interface. The interface components are illustrated in Figure \ref{fig:interface}.

\subsubsection{Intuitive Selection Tools}
\begin{figure}[h]
  \centering
  \includegraphics[width=0.9\linewidth]{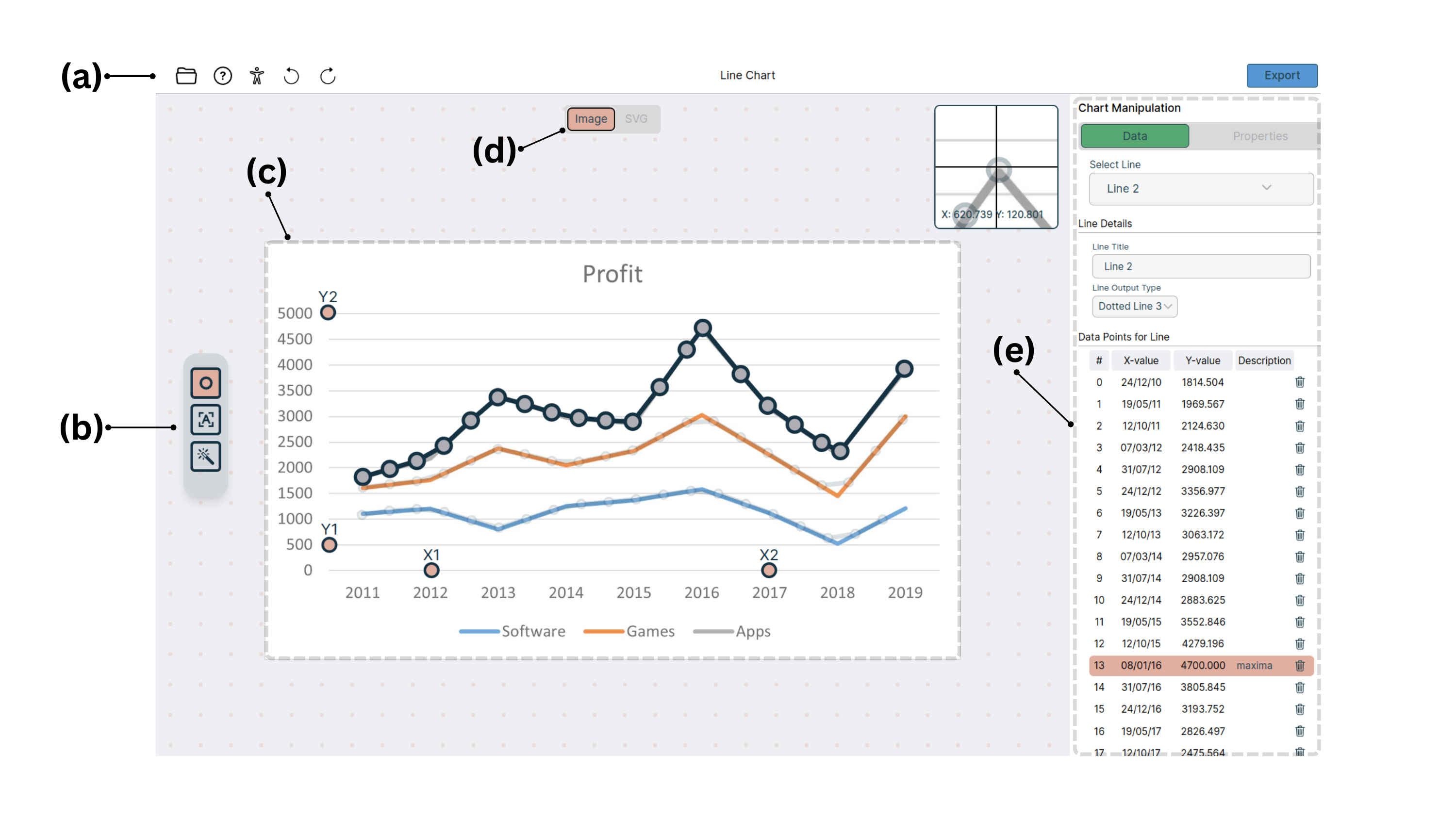}
  \caption{An overview of Chart4Blind interface sections: (a) Home menu for actions like upload, undo, redo, and tutorials. (b) AI toolbar with OCR and segmentation models. (c) Canvas for the uploaded chart, allowing interaction for calibration points and predicted line adjustments. (d) Rendering Module for real-time SVG visualization before export. (e) Metadata section for visualizing extracted line data and seamlessly drag-and-drop of textual content.}
  \Description{An overview of Chart4Blind interface sections: (a) Home menu for actions like upload, undo, redo, and tutorials. (b) AI toolbar with OCR and segmentation models. (c) Canvas for the uploaded chart, allowing interaction for calibration points and predicted line adjustments. (d) Rendering Module for real-time SVG visualization before export. (e) Metadata section for visualizing extracted line data and seamlessly drag-and-drop of textual content.}
  \label{fig:interface}
\end{figure}

In line with our objective to minimize the number of clicks required in the process of chart conversion, we provide two options for text extraction. The first option enables users to one-click on the OCR icon from the toolbar for text extraction, prompting the localization and prediction of all textual content in the chart, Figure \ref{fig:interface} (b). Users can then drag and drop the text into the appropriate fields. The second option allows users to select a specific text of interest. Both options support batch selection for axes labels, which are automatically sorted in the corresponding fields, eliminating the need to select each label separately.

Segmented points on each line are also modifiable. Users have the capability to shift a point, delete, or add points with right and left mouse clicks respectively. To improve accuracy, we incorporated a zoom window that enlarges the surrounding area around the mouse cursor following the PlotDigitizer approach \cite{Huwaldt2023}. Given the potential for errors during this stage, we log a history of labeling actions, enabling users to undo or redo actions within the interface.

\subsubsection{Interactive Tutorials and Accessibility Guidelines}
As we expect users who may not be familiar with chart conversion or accessibility guidelines to use the app, we have incorporated two tutorials. These tutorials serve as a reference to help users understand how to utilize the tool effectively. The first tutorial is interactive and is presented when opening the app for the first time. A sample chart is loaded, with pop-up short animated GIFs and text descriptions. users can interact with tutorials and experiment with the tools at the same time. 

The second tutorial is designed to inform users about the accessibility guidelines. This tutorial showcases an accessible SVG chart within the rendering module. Similar to the previous tutorial, components are highlighted with corresponding descriptions. Notably, the accessibility tutorial has played a crucial role in educating users about the importance of chart descriptions, leading to an increase in the number of charts with accessible descriptions, as will be discussed in our user study.

%% file: chapters/userstudy.tex
\section{Usability Study}
We conducted a user study to see if the current implementation of Chart4Blind fulfils our design principles in terms of supporting user requirements discussed in the findings section \ref{findings} to semi-automatically support the creation of accessible charts according to current standards~\cite{BANA2010}. We also conducted a follow-up study measuring the accessibility of our exports involving BVI individuals. 
\subsection{Participant}
We invited a total of 10 sighted participants (T1-10, age range 22-34 years). Similar to the previous study, we collaborated with ACCESS@KIT to recruit participants via email lists. They did not receive any compensation. We screened participants for basic knowledge of charts: all participants were familiar with reading line charts and frequently worked with them. While their educational backgrounds varied, none of the participants had prior experience with chart accessibility or the conversion process. 
The study was part of a series of studies which were approved by the ethical review committee of Karlsruhe Institute of Technology.
\subsection{Study Design}
We prepared three line charts from our test set, for the conversion process and one simple chart for the tutorial session to help users become familiar with the tool. We consider three levels of complexity of charts to experiment with the conversion process, which we define as follows:

\begin{itemize}
    \item Simple: Few lines with few data points and labels.
    \item Compound: Two or more lines with different label formats.
    \item Dense: Complex lines such as long sinusoidal waves or overlapping trends with text annotations.
\end{itemize}

\begin{figure}[h]
  \centering
  \includegraphics[width=0.7\linewidth]{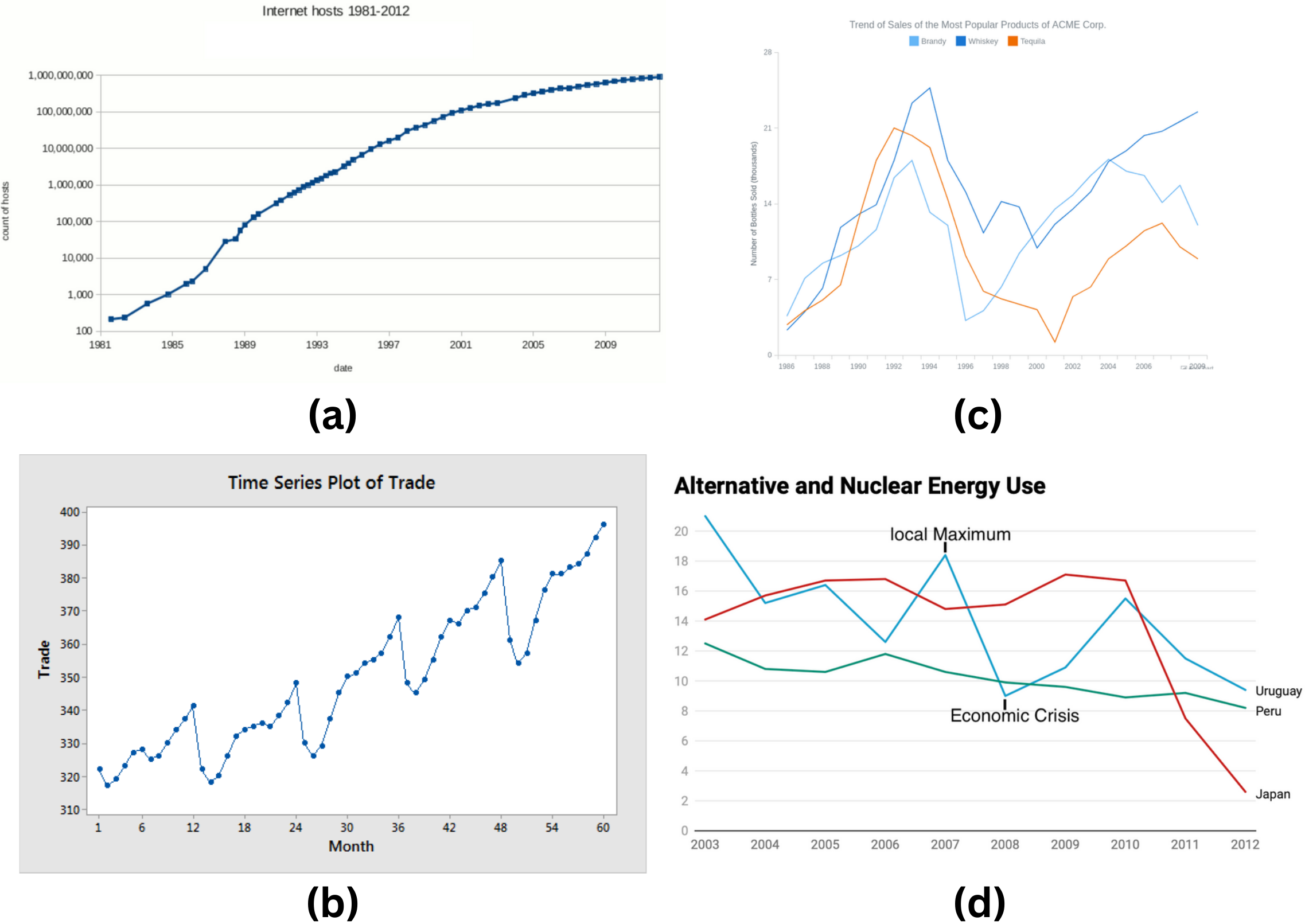}
  \caption{Four line charts with different complexities utilized for the user study: Simple charts (a) and (b) each contain one simple line trend for the tutorial and main session respectively. (c) A compound chart with additional lines overlapped, and visible axes. (d) A dense chart featuring relatively complex trends, point annotations, and less visible axes.}
  \Description{Four line charts with different complexities utilized for the user study: Simple charts (a) and (b) each contain one simple line trend. (c) A compound chart with additional lines overlapped, and visible axes. (d) A dense chart featuring relatively complex trends, point annotations, and less visible axes.}
  \label{fig:userstudyfigures}
\end{figure}

We randomly selected line charts meeting the established criteria to ensure diversity (see Figure \ref{fig:userstudyfigures}). These charts were collected from the recent \textit{LG} dataset \cite{moured2023line}, featuring real charts from public documents across 5 distinct fields (e.g., social science, natural science, etc.). The charts are provided in PNG format, accompanied by the ground-truth hierarchical segmentation masks for all visual and textual elements. 

Each participant completed a total of 3 sessions, progressing from Simpler to Denser charts. For each session, we recorded task completion time (measured in seconds), mouse clicks, SVG, and CSV exports. Heatmaps were generated using the recorded mouse clicks. Line point quality was measured using the Frechet Distance \cite{alt1995computing}, and the SVGs were utilized in the follow-up study, as discussed later.

\subsection{Procedure}
After completing the consent form, participants were invited to a face-to-face user study. They went through a tutorial session to familiarize themselves with the Chart4Blind tool. During the tutorial, we presented the Chart4Blind tool using a sample chart (Figure \ref{fig:userstudyfigures} (a)) that was not used in the subsequent sessions. Additionally, participants were guided to an accessibility tutorial to understand the expected results.  In order to get comfortable with the tool, we allowed the participant to interact with it for 15-20 minutes.

In the main sessions, each of the 10 participants was provided with three charts (see Figure \ref{fig:userstudyfigures} (c-d)) and asked to upload them and start the conversion process. Participants were informed that they were free to choose their preferred approach, whether utilizing the integrated AI tools or performing manual metadata labelling. They were also informed that they could revisit both the tool and accessibility tutorials at any time without losing progress if they had any questions.

After completing all the trials, the participants were presented with the SUS survey \cite{jordan1996usability}, followed by open-ended questions to express their opinions and thoughts. The entire experiment took approximately 90 minutes to complete. We developed an offline experimental system to facilitate the study, setting up Chart4Blind on a local laptop, with the tool accessible through the Chrome browser.  

\subsection{Results and Discussion}
\subsubsection{Task Performance} 
We measured the average time in seconds spent on each chart as illustrated in Figure \ref{fig:userstudyfigures}. For the initial tutorial part, users took, on average, 9min and 5sec (STD: 3min and 45sec), with differences among users—some were interested in understanding more about accessibility guidelines, while others chose to go through the tutorial via GIF animations only.

For the main session, we observed that participants completed the conversion process in an average of 4min and 36sec (STD: 2min 55sec).
We informally asked the most experienced participant about how long it takes on average to create a chart with the tools she used (LibreOffic Draw). She reported a 10 minutes duration.

\begin{figure}[!h]
  \centering
  \includegraphics[width=0.9\linewidth]{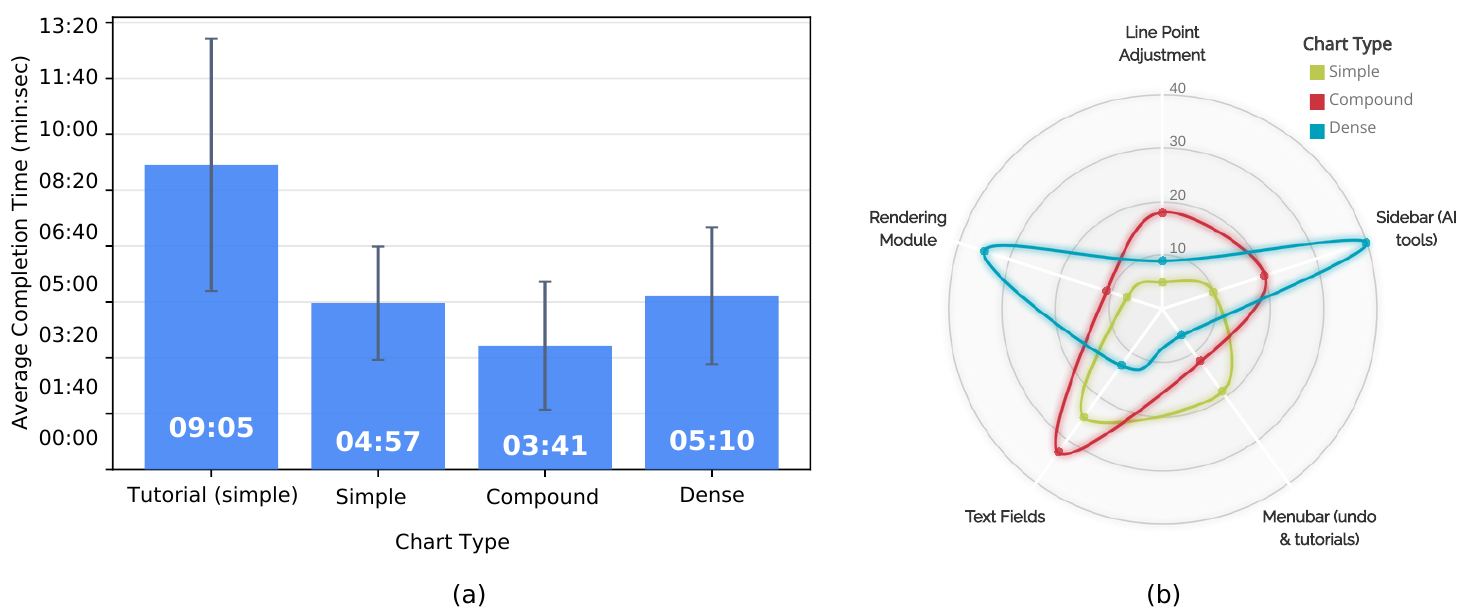}
  \caption{On the left, average conversion task completion time in minutes:seconds. On the right, a radar chart depicting the number of clicks for the top 5 sections interacted with in Chart4Blinds.}
  \Description{On the left, average conversion task completion time in minutes:seconds. On the right, a radar chart depicting the number of clicks for the top 5 sections interacted with in Chart4Blinds.}
  \label{fig:userstudyaverage}
\end{figure}

\subsubsection{Interaction Patterns}
To conduct a more comprehensive analysis of the time spent, we closely examined user click patterns and the duration they spent on various parts of the interface. In Figure \ref{fig:userstudyfigures}-b, a radar chart demonstrates the average number of clicks on the top 5 interacted components of the UI. The analysis suggests an influence on completion time when incorporating the segmentation model. All users started with initial predictions, later making slight adjustments to the misaligned points, specifically focusing on points situated around the corners of the curved lines (see Figure \ref{fig:userstudyfigures}-c). This trend indicates a decline in segmentation performance in these specific areas, suggesting a need for improvement in future work. However, in contrast to annotating the entire line manually, this discrepancy in performance is minimal compared to the overall number of points. In the case of the Dense chart, users edited $10.0\%$ of the overall points.

Further investigation revealed that the time spent varied based on the elements within the charts. For instance, in charts with more textual components like the Compound chart, users tended to spend more time engaging with the text fields. On the other side, in the Dense chart, users engaged more with the rendering module, likely due to a few misaligned points caused by the curved lines. As users gained more experience and knowledge of accessibility guidelines, we observed a decrease in interactions with the menubar in each subsequent session.
\subsubsection{Post-study Feedback}
User satisfaction was assessed through a SUS survey \cite{jordan1996usability} conducted after the interview. The study yielded an average score of $90$. All participants either agreed or strongly agreed that the Chart4Blind tool is valuable for converting line charts into accessible formats and helps to understand the accessibility guidelines. Additionally, they agreed that they would recommend this tool to other colleagues.

We conducted a follow-up discussion to gather participant insights, covering the following aspects:
\begin{itemize}
    \item Describe your experience with the tool.
    \item What did you find most intuitive and why?
    \item What parts do you find difficult or confusing and why?
    \item What suggestions or functions you would like to see or improve in the tool?
\end{itemize}

All participants expressed a positive experience with the tutorials, finding them informative enough to create an accessible chart. They felt confident following the conversion process through the rendering view, even without prior experience. T8 remarked, "I didn't feel the difference in converting Dense and Simple charts; they both took similar effort," indicating that the tool was seamless to utilize regardless of chart complexity.

6 participants found the line segmentation feature valuable as it significantly reduced the effort required to trace lines manually. A few participants also valued the OCR and drag-and-drop features equally alongside the segmentation model.

Interviews further revealed that while Chart4Blind generally met their needs, some participants suggested the following improvements: Three participants mentioned that automating the calibration step would be helpful, expressing an occasional lack of confidence when placing calibration points. T5 explained, "This chart [Dense] has a transparent axis with light gridlines; maybe other charts have no axis. I can't locate efficiently even with the zoom window."

Our sighted participants found the template-based summaries good but expressed a preference for more contextual summarizations reflecting domain knowledge presented in the chart. T10 suggested further integration of natural language models to enhance chart summarization.


%% file: chapters/evaluatingoutputaccessibility.tex
\section{Exploratory Evaluation of Output Accessibility}
\begin{figure}[!h]
  \centering
  \includegraphics[width=0.9\linewidth]{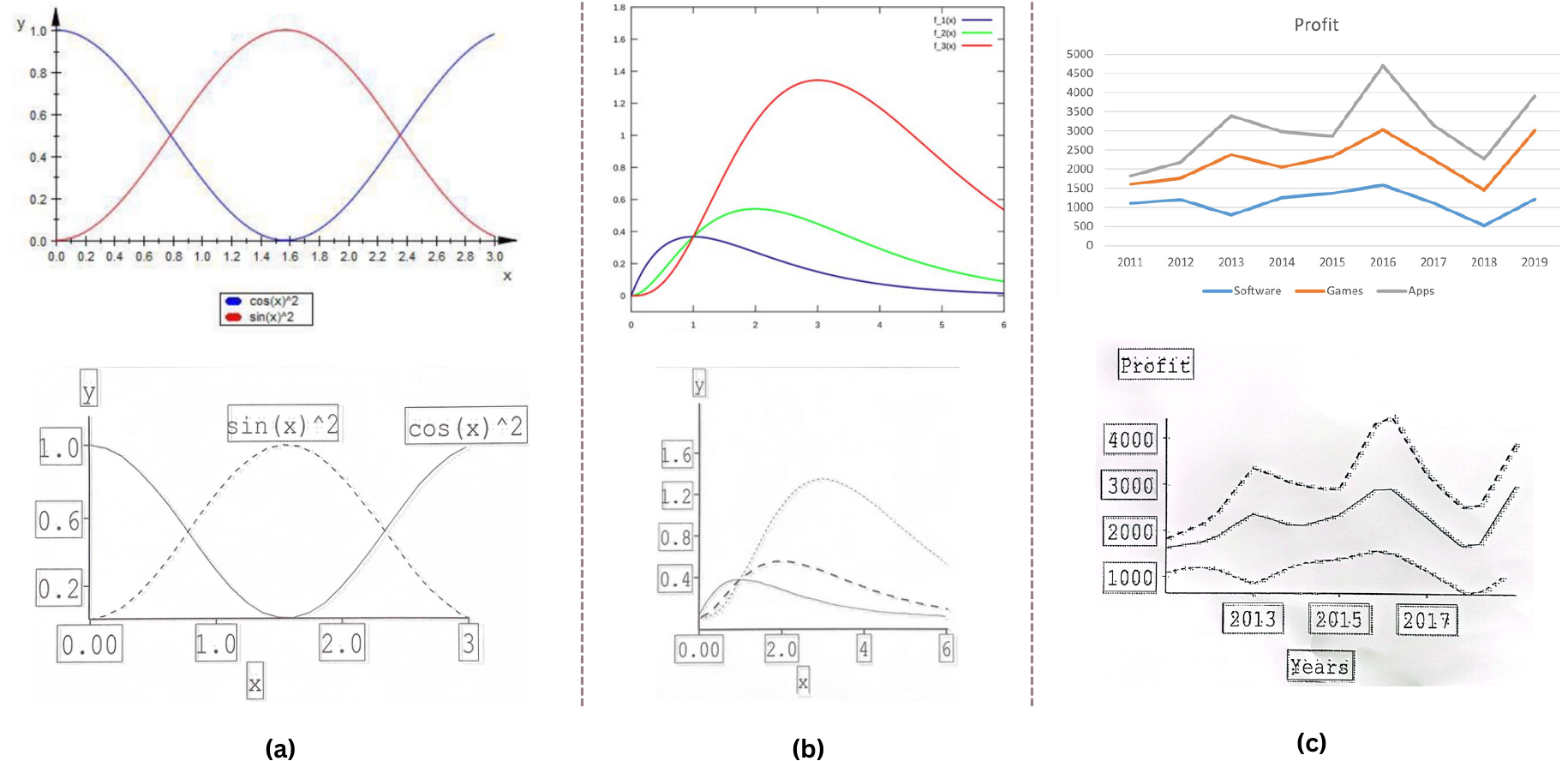}
  \caption{Three printed tactile charts sent to our BVI individuals. The top row displays the original chart images, while the bottom row presents the tactile versions. For better resolution please refer to our supplementary materials.}
  \Description{Three printed tactile charts sent to our BVI individuals. The top row displays the original chart images, while the bottom row presents the tactile versions. For better resolution please refer to our supplementary materials.}
  \label{fig:tactilecharts}
\end{figure}

As an exploratory effort to evaluate output accessibility with assistive technologies, we conducted a remote study. We recruited three visually impaired individuals via the ACCESS@KIT center email list. We randomly selected three converted charts from our test set, shown in Figure \ref{fig:tactilecharts}, and sent them to the participants, including embossed prints via mail and SVG exports via email. Please refer to our supplementary materials for the SVGs. Participants were instructed to review the charts using a screen reader first and then explore the corresponding embossed print. We utilized the SoSciSurvey tool~\cite{Sosci} to create an online questionnaire accessible with screen readers. The complete version of the survey is available in the supplementary materials. Participants were guided to answer two questions: 

\begin{itemize}
    \item Could you describe your experience in using a screen reader to access the provided SVG? Were both the SVG and the accompanying chart description digitally accessible to you?

    \item Please discuss your interaction with the provided embossed print. Were you able to access both the visual and textual elements effectively?

\end{itemize}

All participants are familiar with the usage of screen readers, and with reading charts on tactile prints provided by ACCESS@KIT due to their studies in STEM subjects.

A few notable points were analyzed: two participants found the template summarization informative, while one preferred more comprehensive descriptions that also summarize the overall trends and patterns of the data. Furthermore, participants reported a few missing elements: plot title in chart (a), as well as missing legends in charts (b) and (c), highlighting potential errors made by sighted users when converting the graphics to an accessible format. Regarding the tactile prints, one participant mentioned that a few labels were very close to the border of the text bounding box (see Figure \ref{fig:tactilecharts}-a and b), making it slightly more difficult to interpret.

In response to these observations, we made updates to the SVG rendering attributes to maintain a larger distance between the border and the inner text (see Figure \ref{fig:tactilecharts}-c). Additionally, we are actively working on implementing a status system that notifies the user if any information is still missing before the export.

%% file: chapters/limitationsandfuturework.tex
\section{Limitations and Future Work}
The evaluation of Chart4Blind indicates that utilizing intelligent solutions for chart accessibility enhanced the efficiency and effectiveness of the conversion process for sighted individuals. The study also revealed important lessons and helped us identify the limitations that we would like to lift in future work.

\textbf{Chart analysis limitations:} Currently, Chart4Blind exclusively supports line charts. We initially chose this type because it is commonly found in documents and, unfortunately, due to dataset constraints, we could only train the segmentation model on the LG dataset provided by Omar et al. \cite{moured2023line}. We considered synthesizing charts as an alternative, but it yielded biased and unrealistic performance with the real charts. As the field of chart analysis advances, we aim to support a broader range of chart types, including bar charts, pie charts, and scatter plots. Nevertheless, the current instance segmentation paradigm may not be adaptable to all chart types. For example, segmentation models may struggle with very faint dots in scatter plots, as they consist of very few pixels, potentially increasing the complexity of the models. Future research may introduce additional modalities that users can select from to extract metadata.

\textbf{LLMs for chart summarization:} Large Language Models (LLMs) have recently demonstrated significant potential in achieving contextual understanding from images. For example, Smith et al. highlighted the effectiveness of ByT5 model in generating structured chart summarizations from images for BVI individuals \cite{tang2023vistext}. Our user studies revealed that BVI participants favored having more detailed descriptions of the trends and patterns observed in the chart, while sighted participants aimed to minimize the time spent on generating these descriptions. To address this, we are exploring more collaborative authoring techniques using LLMs to streamline the chart description generation process

\textbf{Limitation of user study:} Our initial evaluation of Chart4Blind was conducted through user studies in a controlled environment, focusing solely on line charts and within a restricted timeframe. Additionally, our interaction logs and collected data are limited, affecting the depth of our analysis. To enable more thorough analysis and contributions, we have made our system and models publicly available, encouraging further exploration from both the accessibility and chart analysis communities. We are currently conducting additional studies with experts to address the aforementioned points.

%% file: chapters/conclusion.tex
\section{Conclusion}
We have introduced Chart4Blind, an interactive web-based tool designed to facilitate a semi-automated accessible chart conversion process using instance segmentation and OCR models. The tool guides inexperienced users in processing online charts and exporting accessible visualizations for blind and visually impaired individuals, supporting various assistive modalities such as embossed papers, screen readers, and tactile displays. We equip the digital exports with chart descriptions generated through template-based summaries. Our tool incorporates interactive features for text selection and annotating line points. A user study involving inexperienced users confirms that this semi-supervised approach, coupled with interactive features, simplifies the task of chart conversion more efficiently and effectively. We anticipate that Chart4Blind will significantly contribute to making information visualization and analytics more accessible to everyone. Future work includes the incorporation of other diagram types. To encourage further research in this relatively unexplored area, Chart4Blind is accessible through the project page \href{https://moured.github.io/chart4blind/}{https://moured.github.io/chart4blind/}.